\documentclass[journal]{IEEEtran}
\ifCLASSINFOpdf
\else
\fi
\usepackage{caption}
\usepackage{graphicx}
\begin{document}
%
\title{Analysis of an Induction Machine Fed by a Space Vector Modulated VSI}
%
%
%

\author{FNU Nishanth~\IEEEmembership{}\\
Dept. of Electrical and Computer Engineering \\University of Wisconsin-Madison\\Madison, WI, USA\\ e-mail: nishanth@wisc.edu}

%
%

\markboth{\emph{}}%
{Shell \MakeLowerecase{\textit{et al.}}: Bare Demo of IEEEtran.cls for IEEE Journals}
%



\maketitle

\begin{abstract}
This paper describes the analysis and simulation performed on a three phase squirrel cage induction motor, fed by a Voltage Source Inverter (VSI), modulated using the space vector modulation scheme. The study mainly focused on analysing the motor performance when fed by a space vector modulated voltage source inverter. The effect of switching frequency on the motor current ripple as well as the frequency spectra of motor current, voltage and torque were studied. The transient response of the inverter-fed motor for a sudden load change was studied and compared to that of a motor fed with an ideal AC source. Results from these studies performed as computer simulations are presented and analysed.
\end{abstract}

\begin{IEEEkeywords}
Induction motor, Space Vector Pulse Width Modulation (SVPWM), Current ripple, Frequency spectra, Transient load change, Switching frequency effects
\end{IEEEkeywords}

%
\IEEEpeerreviewmaketitle

\section{Introduction}
%
%
%
%
\IEEEPARstart{E}{lectric} machines are widely used in industrial drives, motion control and electrified transportation, e.g.:~\cite{175238, 9595085, 9161495,9813839,9826430,9235731}. The squirrel cage induction machine is one of the most widely used electric machines owing to its robustness and reliability. While, the machine can be directly used on-line for constant speed applications, not requiring precision control of speed, it becomes essential to use a power electronic drive if a precise control of the motor speed is desired. The simplest and most popular approach for speed control is the volt/hertz control, where the input voltage magnitude and frequency are varied in a manner as to keep the volt/hertz ratio constant. However, even this requires that a power electronic drive be used to supply the machine with required voltage and the frequency determined for a certain speed command~\cite{simplesvpwm}.
There are other advanced methods available too, all of which require that a power electronic drive be used, thereby making the power electronic drive an important component in the overall electro-mechanical conversion system. A typical power electronic drive comprises of several subsystems and components of which the inverter is of high importance owing to it's role in supplying the electric machine with the required voltage or current waveform. Among the various classes of inverters available, the Voltage Source Inverter (VSI) with Space Vector Pulse Width Modulation (SVPWM) is very popular due to its numerous advantages. However, the use of a VSI and selection of certain parameters for the PWM scheme have an impact on the motor performance. Hence, in the course of this project, the impact of some of the SVPWM parameters on the motor performance was studied by means of computer simulations. The concept of SVPWM is briefly described in Section \ref{svpwm} while Section \ref{study} describes the study undertaken. The results from the study are demonstrated and discussed in Sections \ref{simulation} and \ref{discussion} respectively.
\section{Space Vector Modulation}\label{svpwm}
\begin{figure}
    \centering
    \includegraphics[scale=0.4]{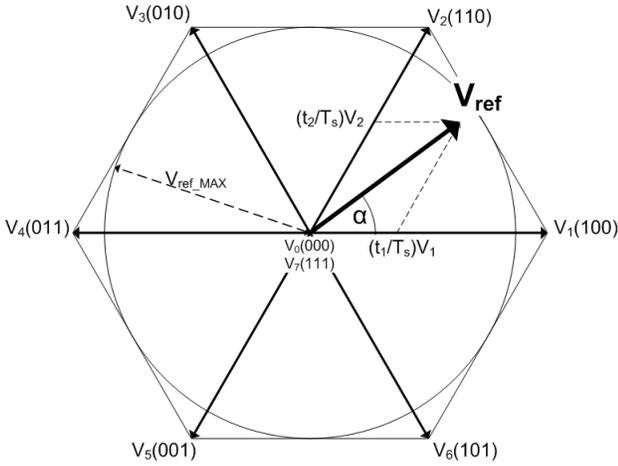}
    \caption{Space vector Diagram}
    \label{fig:SVPWM}
\end{figure}

\begin{figure}
    \centering
    \includegraphics[scale=0.45]{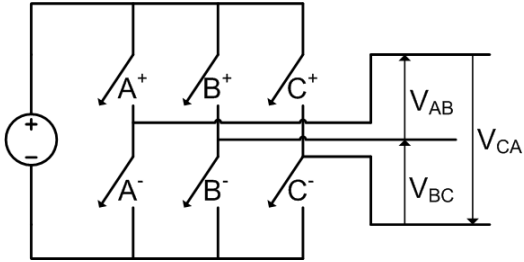}
    \caption{Three Phase 3 Leg VSI}
    \label{fig:Inverter}
\end{figure}

The Space Vector Pulse Width Modulation (SVPWM) is a popular technique used in modulation of VSI  for AC motor drives. The SVPWM implemented in this study was based on the material presented in \cite{simplesvpwm} \cite{bkbose}\cite{lipo_pwm} which is briefly described in this section.
Any set of time dependent three phase waveforms, can be represented in the space phasor notation using the expression 
\begin{equation}
    V_{qds} = V_q - jV_d
\end{equation}
where, $j$ is the complex operator. This is well-known and extensively used in analysis of AC machines. The space vector modulation technique is extension of the same concept to effectively modulate power electronic inverters.
The reference voltages to a three phase inverter can be compactly denoted as $V_{abc}$ These can be transformed to the $d-q$ two phase reference frame (also called $\alpha\beta$ frame) using the sets of equations
\begin{equation}
    V_{qs} = \frac{1}{3}(2V_a-V_b-V_c)
\end{equation}
\begin{equation}
     V_{ds} = \frac{-1}{\sqrt{3}}(V_b-V_c)
\end{equation}
These can be viewed as two components of a space phasor (also called space vector) rotating in space. The magnitude is given by $V_{mag} = \sqrt{V_{ds}^2+V_{qs}^2}$ an the phase angle it makes with the positive real axis is given by $\alpha = tan^{-1}(\frac{-Vd}{Vq})$. The values of $\alpha$ and $V_{mag}$ uniquely determine the position of the space vector at any instant of time. It can be shown that the space vector derived from a balanced set $V_{abc}$ traces a trajectory in space, which is a hexagon. The hexagon is divided into six sectors as shown in Fig.~\ref{fig:SVPWM}. The eight voltage vectors numbered $V_0$ through $V_7$ on the vertices of the hexagon and at the center point are the switching vectors. The 3 phase, 3 leg VSI, modulated using the SVPWM approach is shown in Fig.~\ref{fig:Inverter}. The numbers adjacent to the switching vectors in Fig.~\ref{fig:SVPWM} indicate the states of the Phase A, B and C switches respectively, when the particular switching vector is selected. 

\subsection{Working Principle}
\begin{itemize}
\item The complex space phasor to be applied, $V_{ref}$ is first determined. 
\item The angle $\alpha$ and magnitude $V_{mag}$ are computed. 
\item Based on $\alpha$, the sector of the hexagon, in which the space phasor lies is determined. 
\item Once the sector is identified, the vector $V_ref$ is created using the switching vectors bounding the identified sector and the zero vector, for required dwell times.
\end{itemize}
Each switching period $(T_s)$ is divided into two halves. Consider for example, $V_{ref}$ is in Sector 1. Then, the switching vectors $V_1, V_2, V_0$ and / or $V_7$ are considered. If $V_1$ is applied for a duration $T_1$ and $V_2$ for a duration $T_2$, the zero vectors $V_0$ and/or $V_7$ are applied for $\frac{T_s}{2}-T_1-T_2$. 
The switching durations (also called dwell times) $T_1$ and $T_2$ are computed using the material presented in \cite{lipo_pwm} and summarized in Eqns. \ref{T1} and \ref{T2}
\begin{equation}\label{T1}
T_1 = \frac{V_{ref}sin(\frac{\pi}{3}-\theta_o)}{V_msin(\frac{\pi}{3})}\frac{T_s}{2}
\end{equation}
\begin{equation}\label{T2}
T_2 = \frac{V_{ref}sin(\theta_o)}{V_msin(\frac{\pi}{3})}\frac{T_s}{2}
\end{equation}
where, $\theta_o = \omega_ot$.\\
The same vectors are applied for the same durations for the next half of the switching period, before the vector transitions into the adjacent sector.
The sectors and the vectors that can be used in each sector are summarized in the table~\ref{table1}
\begin{center}
\begin{tabular}{||c || c||}
 \hline
Sector & Vectors\\
 \hline\hline
1 & [1,0,0], [1,1,0], [1,1,1], [0,0,0]\\ 
 \hline
2 & [1,1,0], [0,1,0], [0,0,0], [1,1,1]\\ 
 \hline
3 & [0,1,0], [0,1,1], [1,1,1], [0,0,0]\\
 \hline
4 & [0,1,1], [0,0,1], [0,0,0], [1,1,1]\\
 \hline
5 & [0,0,1], [1,0,1], [1,1,1], [0,0,0]\\
 \hline
6 & [1,0,1], [1,0,0], [0,0,0], [1,1,1]\\
 \hline
 \hline
\end{tabular}
\end{center}
\captionof{table}{Switching Vectors}\label{table1}
The vectors within a sector are not necessarily switched in the order in which they are presented in Table~\ref{table1}. That is, in sector 1 for instance, the switching may be in the order [0,0,0], [1,0,0], [1,1,0], [1,1,1] so that there is just one set of switch state change when transitioning from one vector to another and it would be desirable to minimize the number of switchings in a practical inverter as this would reduce losses associated with switching. There are many more interesting possibilities are presented in \cite{lipo_pwm} about partitioning the zero states and discontinuous SVPWM. Though these were originally proposed to be studied, it wasn't possible to implement them in the given time frame as the basic SVPWM implementation and Machine Model implementation using Matlab-Simulink took most of the effort.
\section{Study Undertaken}\label{study}
The study undertaken concentrated on implementation of a simple SVPWM algorithm and analysis of machine performance when it is fed by an SVPWM inverter. The main components of the study were
\begin{itemize}
    \item Implementation of a simple SVPWM algorithm
    \item Study of frequency spectra for Current, Voltage and Torque
    \item Transient response to a sudden load change
    \item Study of effect of varying switching frequency 
\end{itemize}
The induction motor is a 100 HP, 4 pole, 60Hz machine. It's major parameters as given in the project specification document are summarized in Table \ref{table}.
\begin{center}
\begin{tabular}{||c || c||}
 \hline
Parameter & Rating\\
 \hline\hline
Voltage & 460 Vrms (l-l)\\ 
 \hline
$r_1, r_2$ & 0.0425 $\Omega$\\ 
 \hline
$x_1, x_2$& 0.284 $\Omega$\\
 \hline
$x_m$ & 8.51 $\Omega$\\
 \hline
 $J$ & 2 $kg-m^2$\\
 \hline
 Rated Slip & 0.0177\\
 \hline
 \hline
\end{tabular}
\end{center}
\captionof{table}{Machine Specifications}\label{table}
The analytical model of the machine was derived in the flux-voltage formulation, based on the theory presented in \cite{lipo_book}. As the study focused on understanding the performance of the machine excited by a VSI switched using the SVPWM algorithm, the SVPWM algorithm as described in Section \ref{svpwm} was used. The three main aspects studied were 
\begin{itemize}
    \item Effect of switching frequency on current ripple
    \item Frequency spectra of motor current, voltage and torque under SVPWM inverter excitation 
    \item Transient response to sudden load change
\end{itemize}
The pulse width modulated inverter's output voltage consists of a number of harmonics in addition to the fundamental (desired) voltage. In general, the predominant harmonics occur in the side bands of the fundamental and the switching frequency \cite{mohan_PE}. In case the switching frequency is low, it is evident that that it's sidebands are also low frequency harmonics. As the harmonic amplitude decrease with increase in harmonic number, lower order harmonics have significant amplitude compared to higher order harmonics and are hence undesirable. In addition, the motor acts as a Low-pass filter, analogous to an R-L load. This makes it desirable to push the harmonics to higher frequencies. Hence, higher switching frequency is expected to reduce the current ripple in the machine. 
The frequency spectrum of the machine current, voltage and electromagnetic torque were proposed to be determined from the simulation, so as to assess the components of various harmonics that would be present in them. This in turn would help design suitable filters and would also be useful in the controller design for the machine.
The transient response of the VSI fed induction machine was also proposed to be studied so as to determine the difference between the transient response from that of a machine fed by an ideal AC voltage source.
\section{Simulation Study}\label{simulation}
Simulation study was performed in Matlab Simulink environment. The machine models were derived based on the theory presented in the course and in \cite{lipo_book}. The derived machine model was implemented in Simulink as the author is more familiar with implementing the machine equations in Simulink using the GUI than programming them into matlab for iterative solution. The solver used was ode45 with a step size equal to 0.1 times time period of the switching frequency. All blocks were created based on differential equations characterizing the machine, as discussed in \cite{lipo_book} and in the course. The scope offered by PLECS blockset was used ONLY to plot the frequency spectra, as this was found to be easily done using PLECS as compared to the laborious process in using Matlab.
The Space vector modulation was implemented as Matlab Code and exported into Simulink initially. However, as there were some issues with this approach and it did not work very well, the same was implemented in simulink using basic blocks and worked well. The code and screen shot of simulink implementation adopted in both the approaches are included in the appendix of this report.
\subsection{Initial Conditions and Parameters}
The model was set up such that the machine starts with zero load torque ($T_l = 0)$ impressed on it. At time t = 3.26s, a load torque of 200 Nm is applied to study the response of the machine to a sudden load change. This was done for two models - one fed with an ideal AC voltage source and another fed with the SVPWM inverter. As there were no battery parameters provided in the project instructions, it was assumed that the SVPWM inverter is fed by an ideal DC source. The DC source voltage was set at 625 Volts and the modulation index was set at 0.9 throughout the simulation study to get the rated voltage at the motor stator terminals.
\begin{figure}[h]
    \centering
    \includegraphics[scale=0.3, trim=2cm 1cm 1cm 1cm,clip]{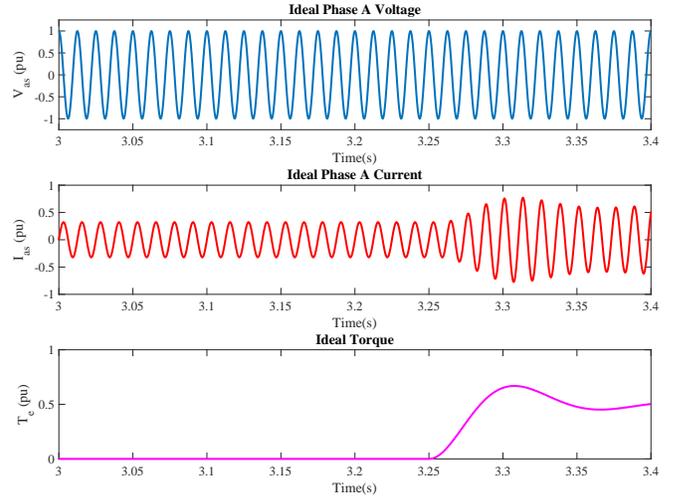}
    \caption{Plots for machine fed by ideal AC source}
    \label{fig:my_label}
\end{figure}
\begin{figure}[h]
    \centering
    \includegraphics[scale=0.3, trim=2cm 1cm 1cm 1cm,clip]{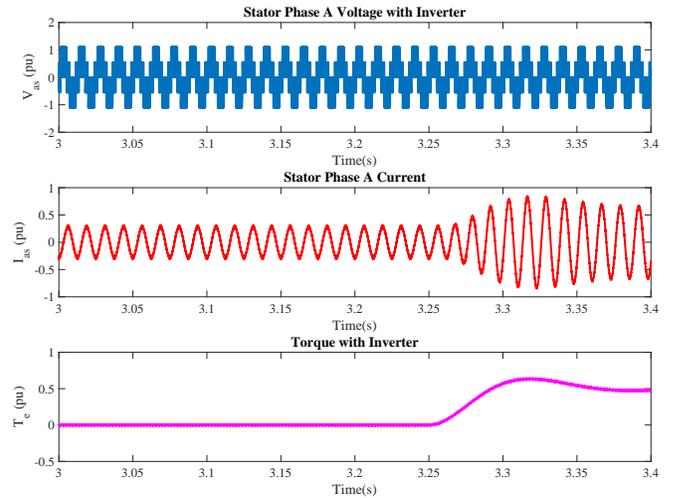}
    \caption{Plots for VSI fed machine at $f_{sw}=7kHz$}
    \label{fig:2}
\end{figure}

\subsection{Effect of switching frequency on current ripple}
\begin{figure}[!h]
    \centering
    \includegraphics[scale=0.3, trim=2cm 1cm 1cm 1cm,clip]{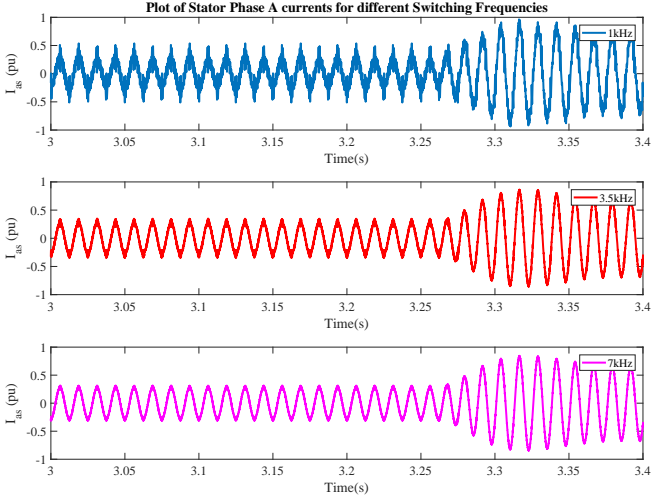}
    \caption{Stator phase A currents for 3 different switching frequencies}
    \label{fig:curr_ripple1}
\end{figure}

\begin{figure}[!h]
    \centering
    \includegraphics[scale=0.3, trim=2cm 1cm 1cm 1cm,clip]{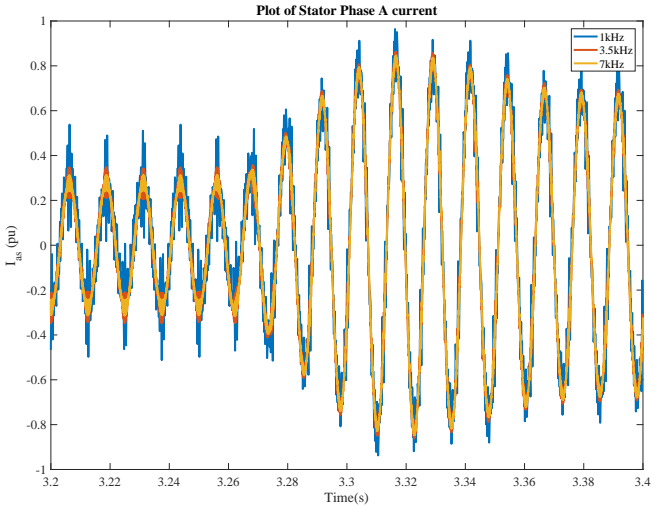}
    \caption{Stator current for different switching frequencies}
    \label{fig:curr_ripple2}
    \end{figure}

The simulation was run for three different switching frequencies namely $1 kHz$, $3.5 kHz$ and $7 kHz$. The Phase A current waveform from these runs are plotted in Fig.~\ref{fig:curr_ripple1} and are plotted for a shorter duration by overlapping over each other, to facilitate comparison, in Fig.~\ref{fig:curr_ripple2}. It is seen from both these plots that the current ripple reduces with increase in switching frequency. This may be easily explained by considering the electric machine as an R-L load, which acts as a low-pass filter. The current ripple mainly arises from the switching and the fact that the voltage is not a pure sinusoid. For an inverter output voltage, as will be seen from the next sub-section, the amplitude of a given harmonic is inversely proportional to it's harmonic order. That is, the higher order harmonics have low amplitudes. However, the harmonics in the side-bands of the switching frequency will have significant amplitude. However, as the switching frequency is increased, their amplitude reduces, contributing to reduction in current ripple at higher switching frequencies. In addition, as the motor current is inductive, it does not immediately fall or rise when a pulse is removed/applied. At lower switching frequencies, the time-duration between PWM pulses is high, allowing the inductive motor current to fall by a reasonable value, resulting in large ripples in the current. However, at the high switching frequencies, this duration between PWM pulses is considerably reduced, leading to reduced ripple.  

\subsection{Frequency Spectra}
\begin{figure}[!h]
    \centering
    \includegraphics[scale=0.3, trim=2cm 0cm 1cm 0cm,clip]{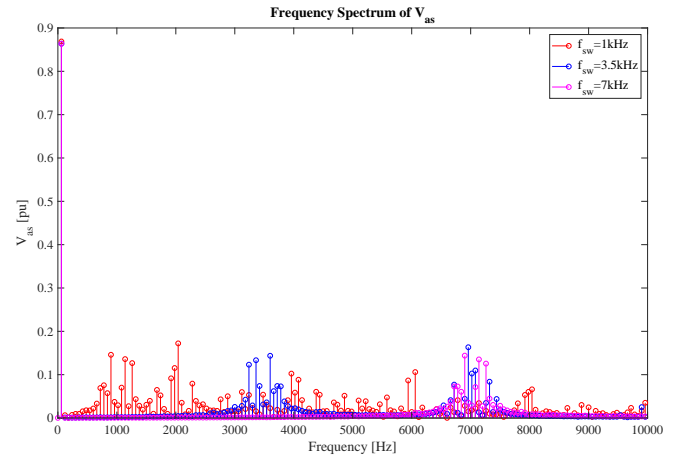}
    \caption{Frequency spectrum of $V_{as}$}
    \label{fig:voltage_spect}
    \end{figure}

\begin{figure}[!h]
    \centering
    \includegraphics[scale=0.3, trim=2cm 1cm 1cm 0cm,clip]{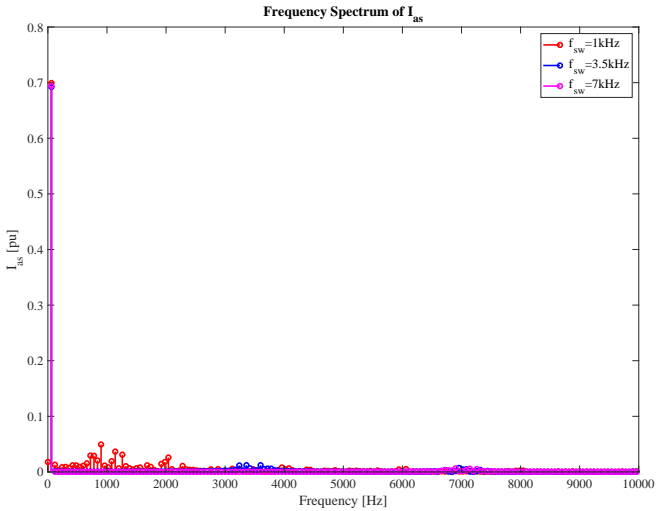}
    \caption{Frequency spectrum of phase A stator current}
    \label{fig:curr_spect}
\end{figure}

\begin{figure}[!h]
    \centering
    \includegraphics[scale=0.35, trim=0cm 0cm 1cm 0cm,clip]{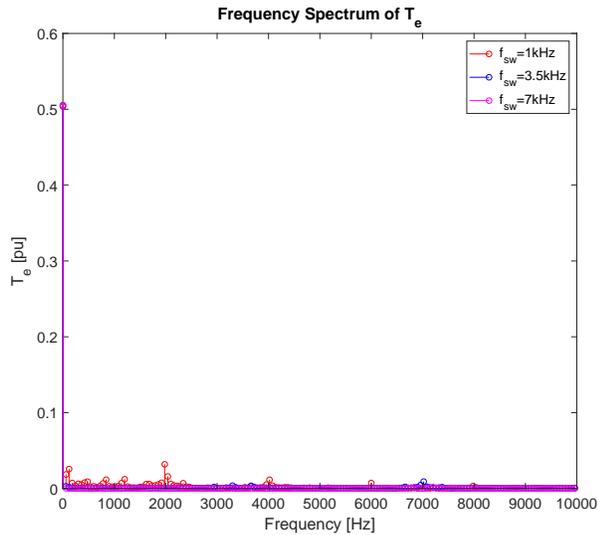}
    \caption{Frequency spectrum of Torque}
    \label{fig:torque_spect}
\end{figure}

Continuing with the analysis performed in the previous sub-section, the frequency spectra of the motor current, applied voltage and torque are plotted for the three switching frequencies of $1kHz$, $3.5kHz$ and $7kHz$. The data was obtained using the PLECS scope from the PLECS blockset in Matlab, saved as a csv file and plotted using Matlab. The base frequency for this was taken as the motor's rated frequency of 60Hz.
It is clearly seen from the voltage spectrum in Fig.~\ref{fig:voltage_spect} that the biggest component is the fundamental and the components along the side band of the switching frequency $f_{sw}$ and their multiples. It is also seen that with the higher switching frequency, many harmonics in the range between $f_o = 60Hz$ and $f_{sw}$, the switching frequency get suppressed. Hence, the use of higher switching frequency is desirable, provided the increase in switching losses that occur with increasing the switching frequency is not predominant.
Another interesting observation here is the difference in relative harmonic amplitudes between the voltage spectra in Fig.~\ref{fig:voltage_spect} and current spectra in Fig.~\ref{fig:curr_spect}. It is clearly seen that the ratio of amplitude of harmonic current to the fundamental current is clearly lower than the ratio of amplitude of harmonic voltage to the fundamental voltage. That is, for any $n^{th}$ harmonic,
\begin{equation}
\frac{V_n}{V_1} > \frac{I_n}{I_1}    
\end{equation}
This is because the motor, which is similar to an R-L circuit, acts as a low-pass filter and attenuates the current harmonics. 
Generally, Total Harmonic Distortion (THD), is used as a measure of the waveform quality \cite{mohan_PE} and defined by
\begin{equation}
THD = \sqrt{\frac{\sum_{n=2}^{\infty}{f_n}^2}{{f_1}^2}}
\end{equation}
where $f$ is any quantity voltage or current and $n$ is the harmonic order. It can be easily shown that the THD for current $I_{as}$ will be lower than the THD for voltage $V_{as}$. This implies that the current $I_{as}$ is more close to sinusoidal than the voltage $V_{as}$ which is confirmed by Fig.~\ref{fig:2}.
The torque spectra in Fig.~\ref{fig:torque_spect} shows that the Torque is near DC and the higher harmonics are negligible. This is predominantly due to the mechanical inertia of the system, which damps out ripples.
\subsection{Transient response to load change}
\begin{figure}[h]
    \centering
    \includegraphics[scale=0.3, trim=2cm 1cm 1cm 1cm,clip]{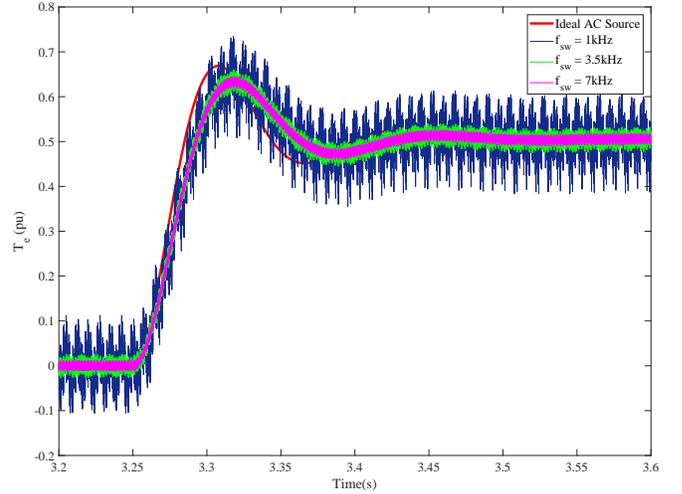}
    \caption{Torque response during transient load change}
    \label{fig:Transient_torque}
    \end{figure}

\begin{figure}[h]
    \centering
    \includegraphics[scale=0.3, trim=2cm 1cm 1cm 1cm,clip]{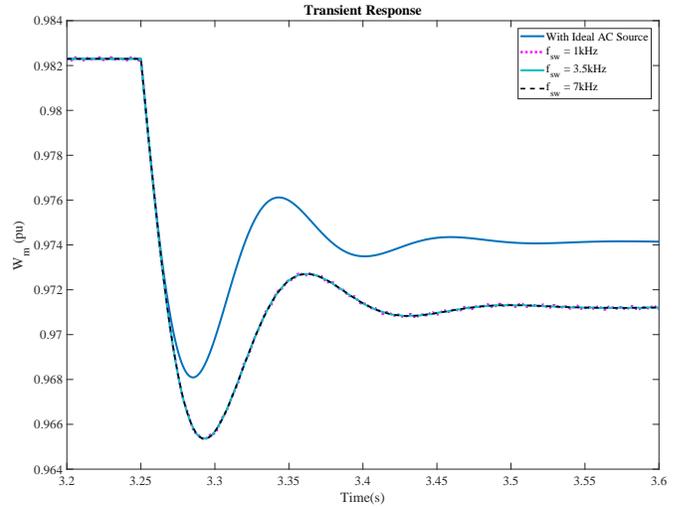}
    \caption{Speed response during transient load change}
    \label{fig:Transient_speed}
    \end{figure}

At t = 3.26s, a load of 200Nm (0.50 pu) is applied on the machine to study the response to load change. The change in current is seen in Fig.~\ref{fig:curr_ripple2} while the torque and speed waveforms are shown in Fig.~\ref{fig:Transient_torque} and Fig.~\ref{fig:Transient_speed} respectively.

It is seen from Fig.~\ref{fig:Transient_speed} that when 0.5 pu  of load torque is suddenly applied, the speed dips during the transient condition. This dip in more pronounced for the inverter fed machine than for the machine fed with the ideal AC source. The difference in speed dip is about 0.0025pu or about 4 $rev/min$. That is, the speed of the inverter fed machine, dips by a larger amount. Also, it is interesting to observe that the mechanical speed that the machines settle to is not the same. While the machine fed with ideal AC source settles to about 0.974pu speed, the inverter fed machines settle to about 0.971pu. While this again, is not a huge difference, it may be due to the fact that the fundamental component of current (at 60 Hz), which is responsible for torque production, is slightly lower in the inverter fed machine, as a result of which, the machine has a greater slip, in order to meet the increased torque demand. This may possibly be resolved by having a higher DC bus and operating the inverter at a high amplitude modulation index, so as to supply the required fundamental current. Though it may not be apparent at first from the Fig~\ref{fig:Transient_speed}, it is seen on closer observation that the inverter fed machines have a tiny ripple on their speed too, which again decreases with increasing switching frequency. This ripple is not as large as the torque ripples seen in Fig.~\ref{fig:Transient_torque}.    

\section{Discussion}\label{discussion}
The analysis performed in this paper, heavily made use of simulation tools. Hence, the accuracy of some of the results would be determined by the step size and numerical approximations made. However, settings in the simulation environment, especially those associated with the step size, were made such that the results are quite accurate. Some of the results observed can be intuitively explained and have been done so when describing them in the previous sections. Analytical solution is generally quite complex for PWM systems. Based on the material presented in \cite{holmes}, it was attempted to try and obtain closed form expression to validate the observations from simulations. While the concepts were easily understandable, the algebra involved was extremely cumbersome. Hence only intutive explanations were presented for the observations.

\section{Conclusion}
This report provided an overview of the study performed on a three phase induction machine fed by SVPWM modulated VSI. It was observed that increased inverter switching frequency reduced the current and torque ripple, reduced the harmonic content and improved the waveform quality, resulting in a lower Total Harmonic Distortion (THD). Based on these observations, it can be concluded that use of higher switching frequency is a better choice for SVPWM based VSI drives. In addition the study on transient response revealed that the inverter fed induction machine while exhibiting similar performance, had a larger dip in the speed during a sudden loading as compared to a machine fed by an ideal AC source. In addition, it was also observed that following the transient condition, the VSI fed motor settled at a speed slightly lower than the Ideal AC source fed machine. This, however may be an issue with the DC bus voltage level and modulation index used for the VSI.

\appendices
\section{List of Programs used}
MathWorks Matlab and Simulink were used for this simulation. One a scope block from Plexim PLECS blockset was used to obtain the frequency spectra. No other software was used. The simulink usage was to create and link different models. \textbf{In-built blocksets/models were NOT used}. Appendix B and C present the Matlab code and Simulink Models developed, in detail. Appendix D presents the code from an initial attempt to implement the SVPWM in Matlab alone, which was not used later owing to a few errors and difficulties with interfacing.

\ifCLASSOPTIONcaptionsoff
  \newpage
\fi
\bibliographystyle{IEEEtran}
\bibliography{references}
\end{document}